\documentclass[10pt,onecolumn]{IEEEtran}

\usepackage{color}

\usepackage{soul}
\usepackage{authblk}
\usepackage{gensymb}
\usepackage[font=footnotesize]{caption}
\renewcommand{\baselinestretch}{.95}
\usepackage{graphicx}
\usepackage{subfigure}

\ifCLASSINFOpdf
\else
\fi

\usepackage[pscoord]{eso-pic}

\usepackage{amsmath, amssymb, amsthm}
\usepackage{array}
\usepackage{color}
\usepackage{subfigure}
\usepackage{tikz}
\usepackage{textcomp}
\usepackage{hyperref}
\usepackage{lipsum}

\newcommand\copyrighttext{%
  \footnotesize \textcolor{blue}{\textcopyright 2019 IEEE. Personal use of this material is permitted.  Permission from IEEE must be obtained for all other uses, in any current or future media, including reprinting/republishing this material for advertising or promotional purposes, creating new collective works, for resale or redistribution to servers or lists, or reuse of any copyrighted component of this work in other works}}
\newcommand\copyrightnotice{%
\begin{tikzpicture}[remember picture,overlay]
\node[anchor=south,yshift=10pt] at (current page.south) {\fbox{\parbox{\dimexpr\textwidth-\fboxsep-\fboxrule\relax}{\copyrighttext}}};
\end{tikzpicture}%
}

\begin{document}

\title{The Adversarial Machine Learning Conundrum: Can The Insecurity of ML Become The  Achilles' Heel of Cognitive Networks?}

\author{Muhammad Usama, Junaid Qadir, Ala Al-Fuqaha, Mounir Hamdi}


\maketitle
\copyrightnotice

\begin{abstract}
The holy grail of networking is to create \textit{cognitive networks} that organize, manage, and drive themselves. Such a vision now seems attainable thanks in large part to the progress in the field of machine learning (ML), which has now already disrupted a number of industries and revolutionized practically all fields of research. But are the ML models foolproof and robust to security attacks to be in charge of managing the network? Unfortunately, many modern ML models are easily misled by simple and easily-crafted adversarial perturbations, which does not bode well for the future of ML-based cognitive networks unless ML vulnerabilities for the cognitive networking environment are identified, addressed, and fixed. The purpose of this article is to highlight the problem of insecure ML and to sensitize the readers to the danger of adversarial ML by showing how an easily-crafted adversarial ML example can compromise the operations of the cognitive self-driving network. In this paper, we demonstrate adversarial attacks on two simple yet representative cognitive networking  applications (namely, intrusion detection and network traffic classification). We also provide some guidelines to design secure ML models for cognitive networks that are robust to adversarial attacks on the ML pipeline of cognitive networks.  
\end{abstract}

\begin{IEEEkeywords}
Cognitive Networks, Self Driving Networks, Security of Machine Learning (ML), Adversarial Machine Learning, Intrusion Detection System (IDS), Network Traffic Classification
\end{IEEEkeywords}

\IEEEpeerreviewmaketitle

\section{Introduction}

\label{sec:Introduction}


The \textit{cognitive networking} idea---a recurring motif in networking research that has been expressed in various guises such as \textit{autonomic networking}, \textit{self-organized networking}, \textit{knowledge-based networking}, and most recently as \textit{self-driving networking} \cite{feamster2017and}. Now, this idea appears to be within grasp thanks to the tremendous strides made in the field machine learning (ML) that have transformed the entire fields of vision, language, speech, and information processing.  Many proponents are optimistic that ML will play a central role in enabling the future self-driving cognitive networks by comprehensively automating the cognitive networking tasks such as (1) real-time telemetry, (2) network automation, (3) network intent modeling, and network decision making.

\begin{figure}[h] 
    \centering    
    \centerline{\includegraphics[width=.75\textwidth]{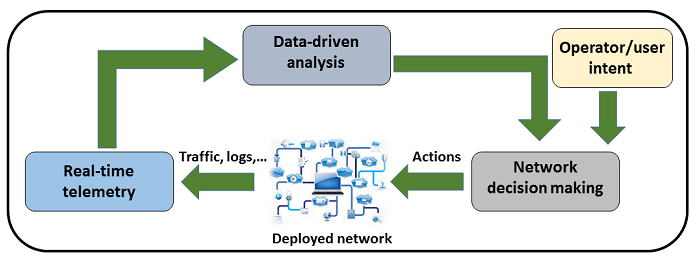}}
    \caption{The cognitive self-driving networking components and their related tasks are highlighted to provide the reader with a basic understanding of how cognitive self-driving networks can be realized. Supervised and reinforcement learning techniques are expected to play a vital role in achieving most of the tasks.}
    \label{fig1}
\end{figure}

A broad illustration of the various tasks involved in the operations of cognitive self-driving networks is provided in Figure \ref{fig1}. Many of these highlighted tasks require a data-driven learning and inference process. Hence, they can benefit from using a ML pipeline involving methods such as (deep) supervised ML and reinforcement learning. 

However, despite the great promise and success of ML methods, the recent discovery of the susceptibility of ML models to security problems has dampened the optimism around the use of ML in cognitive networking. The major reasons for the security vulnerabilities of ML models are the underlying implicit assumption that the training and test data are similar in distribution and that the test examples are benign and not adversarial. 

An ``\textit{adversarial example}'' is defined as an imperceptible minor perturbation of the input that an adversary especially crafts to maximize the prediction error of the ML model \cite{goodfellow2018making}. Deep neural networks (DNNs) in particular have been shown to be very vulnerable to such adversarial examples \cite{papernot2016limitations}. It is worth noting that DNNs are not the only ML models vulnerable to adversarial examples; the problem is much broader and many other ML systems---including reinforcement-learning (RL) and generative models \cite{vorobeychik2018adversarial} are also susceptible to adversarial examples.

\begin{figure}[h]
    \centering    
    \centerline{\includegraphics[width=1.0\textwidth]{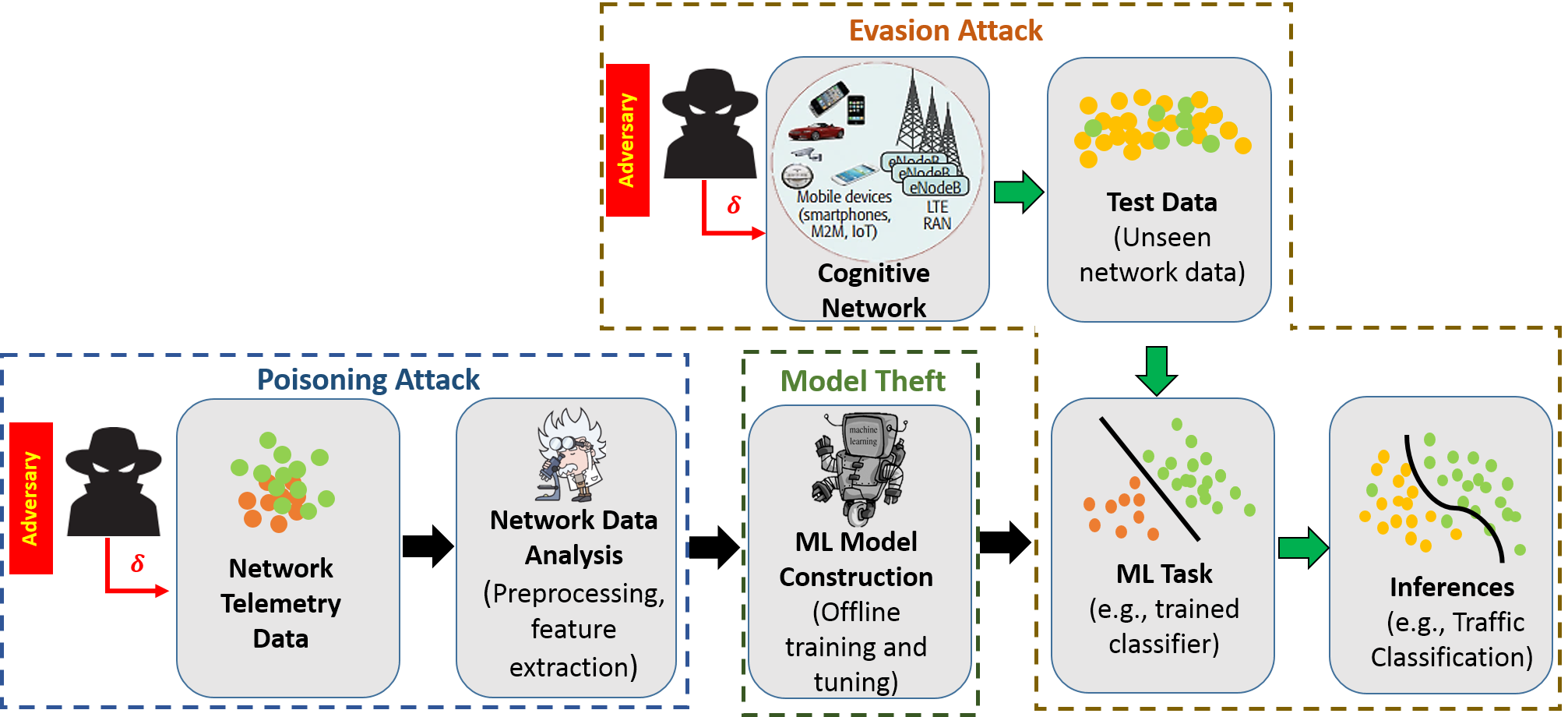}}
    \caption{The ML pipeline in cognitive networking, through which we learn insights from raw network telemetry data by passing it through preprocessing and feature extraction stages and then construct a ML model for some specific task (e.g., intrusion detection). Since our focus is on adversarial ML attacks, we also highlight which parts of the pipeline are vulnerable to poisoning and evasion attacks.}
    \label{fig:pipeline}
\end{figure}

Adversarial ML is now a fast-expanding field attracting significant attention from the industry and academia \cite{vorobeychik2018adversarial}. Although ML vulnerabilities in domains such as vision, image, audio are now well-known, relatively little attention has focused on adversarial attacks on cognitive networking ML models. An illustration of the ML pipeline in cognitive self-driving networks along with potential security attacks that may be launched on its components is depicted next in Figure \ref{fig:pipeline}. 

Due to the rising popularity of cognitive networking, and self-driving networks, ML models used in the context of cognitive networks have become high-profile targets for malevolent adversaries who are interested in compromising the integrity and availability of these ML models. The resulting threat becomes more serious when cognitive networking breaks through into safety-critical networks (such as self-driving vehicles and vehicular networks, Internet of Things, smart city services, and cyber-physical systems); then it will no longer be only computer systems and their accessories that are at risk, but the security of \textit{everything and everyone} will be threatened.

The main \textit{contribution of this paper} is to highlight the vulnerability of ML-based functionality in modern cognitive networks to adversarial attacks and to review the state of the art in the application of adversarial ML techniques in networking. We also provide recommendations for developing robust ML models for self-driving cognitive networks.  In this paper, we only consider the adversarial ML attacks on the network telemetry component of the cognitive self-driving network as other cognitive self-driving networking components rely on the data provided by this critical component. Any fault in the telemetry part will result in less efficient self-driving networks. Since real-time telemetry uses supervised ML schemes, we have chosen anomaly-based intrusion detection and network traffic classification as case studies to highlight the adversarial ML threat on cognitive self-driving networks. Adversarial ML attacks on other components of cognitive self-driving networks which are mostly based on reinforcement learning are left as a future direction. We also propose a concrete novel network-specific adversarial ML attack on an anomaly-based intrusion detection system and network traffic classification to highlight potential issues that may arise when adversarial ML attacks are launched on future ML-based cognitive self-driving networks. For cognitive networking to really take off, it is extremely important that the underlying technology has to be robust to all kinds of potential problems---be they accidental, intentional, or adversarial.

\section{Background and Related Work}
\label{sec:Background}

In this section, we discuss the challenges posed by adversarial ML attacks and then propose a taxonomy of adversarial ML attacks. We will then survey the proposed adversarial attacks and defenses. After that, we highlight the state-of-the-art in adversarial ML attacks on self-driving cognitive networks to emphasize that this area of research needs special attention as networking moves from traditional networking to self-driving cognitive networks.

\subsection{Challenges posed by adversarial ML attacks}

ML adds substantially to the worries of security practitioners by expanding an already-broad attack surface, comprising standard but still potent attacks. In addition, the future densely-connected IoT-era cognitive networking will likely expose new vulnerabilities and a wider attack surface through its emphasis on massive connectivity. Previous research has shown that a motivated well-resourced adversary can arbitrarily alter data and labels to compromise ML models completely and induce an error rate of up to 100\% \cite{biggio2018wild}. 

Another important reason for adversarial ML attacks is the lack of a better understanding of how modern ML frameworks such as DNNs operate. Multiple explanations for the sensitivity of the ML models to adversarial examples have been provided in the literature \cite{vorobeychik2018adversarial} including the nonlinearity of the DNN models, which can assign random labels in areas that are under-explored in the training set. But such a hypothesis fails to explain the transferability\footnote{The \textit{transferability} of an adversarial example refers to the property that adversarial examples produced to mislead a particular ML model can be used to mislead other ML models as well, even if their architectures greatly differ from each other.} of adversarial examples from one ML model to another. In addition, it is not only the nonlinear DNN models that suffer from these attacks, but linear models have also been shown to be vulnerable to adversarial examples \cite{vorobeychik2018adversarial}. While the reasons for the capitulation of the ML models to adversarial examples are still not well-known, it is clear that these adversarial attacks pose a grave danger to the security of future cognitive networks, which requires immediate attention from the community. 


Adversarial examples are especially challenging due to the asymmetric nature of adversarial ML attacks. The asymmetry implies that the job of the defender is to secure the entire attack surface all the time, but the attacker only has to find a single kink in the surface. The attacker also attacks surreptitiously, much like in guerrilla warfare, using previously-unseen attacks at a time of its own choosing. The attacker constructs these attacks creatively by adding adversarial noise to incur the \textit{worst-case domain shifts} in the input in a bid to elicit incorrect results by the ML model. This stacks the odds in favor of the attacker and the battle becomes manifestly unfair one. To prove that the ML model is secure against attacks, the defender has to anticipate the threat models and provide formal proof that demonstrates the fortitude to withstand the assumed threats.

With the well-known attacks proposed in the literature \cite{biggio2018wild}, the bar of effort required for launching new attacks has lowered since the same canned attacks can be used by others. Although Sommer and Paxson \cite{sommer2010outside} were probably right in 2010 to downplay the potential of security attacks on ML saying ``exploiting the specifics of a machine learning implementation requires significant effort, time, and expertise on the attacker's side,'' the danger is real now when an attack can be launched on ML-based implementations with minimal effort, time, and expertise.

\subsection{Taxonomy of Security Attacks on Machine Learning}

\begin{figure*} 
    \centering    
    \centerline{\includegraphics[width=1.0\textwidth]{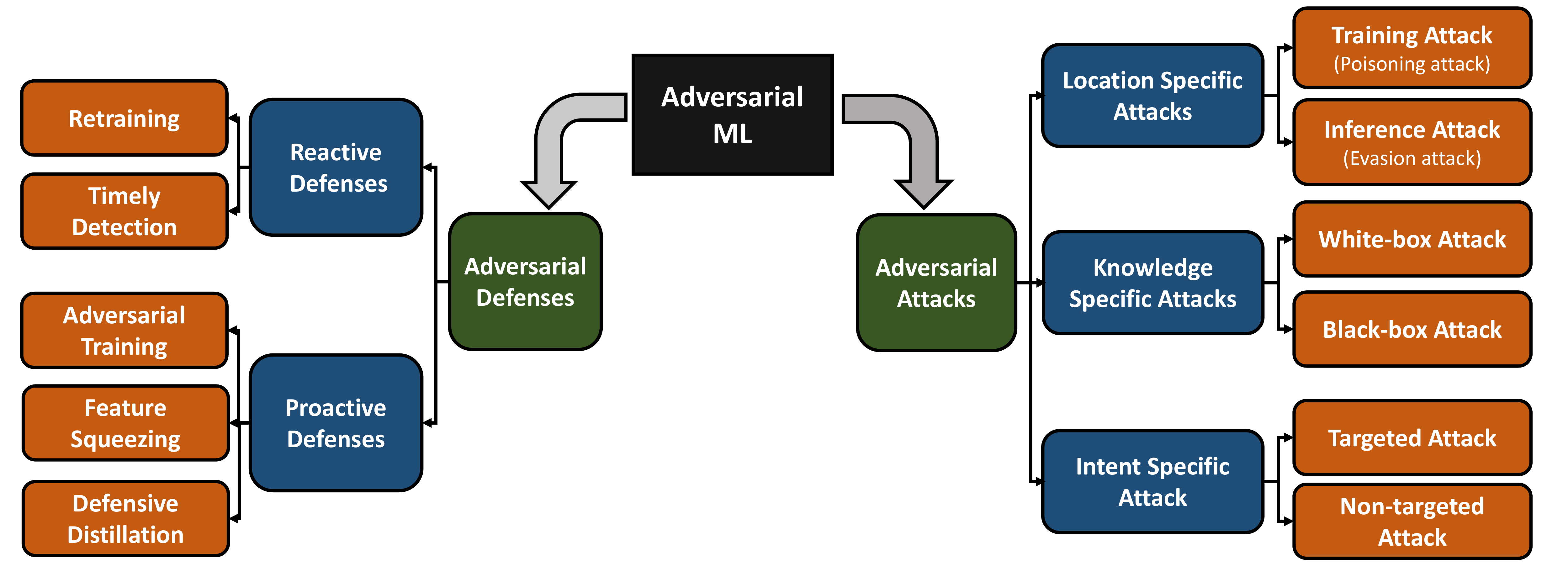}}
    \caption{A taxonomy of adversarial ML attacks is provided in which we sub-classify adversarial attacks and adversarial defense strategies.}
    \label{fig:taxonomy}
\end{figure*}

In this section, we aim to communicate the big picture of adversarial ML security by referring to the ML pipeline for cognitive networking (Figure \ref{fig:pipeline}) and a taxonomy of adversarial ML related issues that we developed (Figure \ref{fig:taxonomy}).

\vspace{1mm}
\textbf{Based on the attack's location on the ML pipeline}, security attacks on ML models can be classified into two categories.  \textit{Firstly}, in a \textit{poisoning} or \textit{training} attack, the attacker can access and adversarially poison the training data in a bid to maximize the classification error. Attacks during the training phase can also include theft of the intellectual property (IP) if the training is outsourced to some external provider. \textit{Secondly}, in an \textit{evasion} or \textit{inference} attack, on the other hand, the attacks attempt to perturb the text/inference input through the creation of adversarial examples to compromise the ML model. The attacks can also attempt to steal the ML model IP through a side channel through successive polling. Other types of attacks may involve obtaining physical access and intrusion into the hardware.

\vspace{1mm}
\textbf{Based on the adversary's knowledge}, adversarial ML attacks can be categorized into two types. In a \textit{white-box attack}, it is assumed that the adversary has perfect knowledge of the ML architecture, training/testing data, and the hyperparameters of the model. In contrast, in a  \textit{black-box attack}, it is considered that the adversary has partial access or no access to the deployed model. Based on the knowledge/access of the adversary, black-box attacks are further divided into two categories; namely, \textit{query-based attacks} and \textit{zero-query attacks}. A black-box attack where an adversary can act as a standard user and query the ML model for a response and later use that query-response pair to generate an adversarial example is known as a query-based attack. The zero-query attack is defined as a black-box attack where the adversary has no access to the deployed ML model but has only a few test samples available to craft adversarial examples against deployed ML model.
\vspace{1mm}

\textbf{Based on the adversarial intent specificity}, we can further divide evasion attacks into two classes. In a \textit{targeted attack}, the attacker aims to fool the ML classifier to classify all adversarial samples in one class by maximizing the probability of the targeted attack. For example, an adversary that wants to disguise the intrusive traffic as normal network traffic can create a perturbation that maximizes the classification probability of the normal traffic class. In a \textit{non-targeted attacks}, the attacker aims to fool the ML classifier by assigning an adversarial sample to any other class except the original one. These attacks are performed by minimizing the probability of the original class that ensures that adversarial sample will not get classified in the original class.

All classification schemes depicted in the taxonomy are directly related to the intent/goal of the adversary. Most of the existing adversarial ML attacks are white-box attacks, which are later converted to black-box attacks by exploiting the transferability property of adversarial examples \cite{szegedy2013intriguing}. The transferability property of adversarial ML means that adversarial perturbations generated for one ML model will often mislead other unseen ML models. Related research has been carried out on adversarial pattern recognition for more than a decade, and even before that there was a smattering of works focused on performing ML in the presence of malicious errors \cite{biggio2018wild}.


\subsection{Brief Review of the Adversarial ML Literature}


\vspace{2mm}
\subsubsection{Adversarial ML Attacks Proposed in Literature} 

ML models, especially those that are DNN-based are very vulnerable to adversarial perturbations. An adversarial sample $x^*$ is created by adding a small carefully crafted perturbation $\delta$ to the correctly classified sample $x$. The perturbation $\delta$ is calculated by approximating the optimization problem given in equation 1 iteratively until the crafted adversarial example gets classified by ML classifier $f(.)$ in targeted class $t$. 

\begin{equation}
    x^* = x + \arg \underset{\delta{_x}}{\text{min}} \{\|\delta\|: f(x + \delta) = t\} 
\end{equation}

In 2013, Szegedy et al. \cite{szegedy2013intriguing} reported that the DNN input-output mapping is fairly discontinuous and DNNs are not robust to small perturbations in the input. This triggered an extraordinary interest in adversarial ML attacks. In 2014, Goodfellow et al. \cite{goodfellow2014explaining} proposed a non-targeted, element-wise adversarial example generation method, where the adversarial perturbation is generated by performing only a single-step gradient update in the direction of the gradient at each element of the input example. This method of generating adversarial ML attacks is called the \textit{fast gradient sign method} (FGSM). Kurakin et al. \cite{usama2018adv} proposed the \textit{basic iterative method} (BIM) attack which improves the FGSM attack by introducing an iterative small perturbation optimization method for generating adversarial examples. Papernot et al. \cite{papernot2016limitations} proposed a targeted saliency map based attack, where saliency map is used in an iterative manner to find the most significant features of the input that when fractionally perturbed cause DNNs to misclassify, this adversarial attack is known as \textit{Jacobian saliency map based attack} (JSMA). Carlini et al. \cite{carlini2017towards} proposed three targeted and iterative adversarial ML attacks by exploiting the three different distance matrices $(L_0, L_2,$ and $L_\infty)$ and highlighted that the defensive distillation method \cite{papernot2016distillation} deployed to increase the robustness of DNNs is not enough for building deterrence against adversarial ML attacks. Most of the adversarial ML attacks are white-box attacks which are later converted to black-box attacks by exploiting the property of transferability of adversarial examples. More details on available adversarial ML attacks and their applications are reviewed in \cite{vorobeychik2018adversarial, corona2013adversarial}.

\vspace{2mm}
\subsubsection{Adversarial ML Defenses Proposed in Literature} 

In response to adversarial ML attacks, researchers have come up with some defenses (some of which focus on detection while others focus on prevention). Generally, defenses against adversarial examples are divided into two broader categories as shown in Figure 3. These categories are \textit{reactive defenses} and \textit{proactive defenses}. Reactive defenses involve retraining or re-configuring the ML model after the adversarial ML attack or timely detection of adversarial attack in order to save critical information. Proactive defenses involve pre-emption of adversarial attacks and preparing the ML model to defend against them. The three major techniques of proactive defenses are \textit{adversarial training, feature squeezing}, and \textit{defensive distillation}.

The technique of \textit{adversarial training} proposed by Goodfellow et al. \cite{goodfellow2014explaining} requires that classifiers be preemptively trained on adversarial perturbations; this defense provides robustness against adversarial examples the classifier is trained on but any perturbation on which the classifier has not been trained can still evade the classifier. Xu et al. \cite{xu2017feature} proposed feature squeezing as another approach for hardening the ML schemes against adversarial attacks. Feature squeezing is a process of reducing the search space available to the adversary by fusing samples that correspond to different feature vectors in the original space into a single sample. Another solution called \textit{network distillation} was proposed by Papernot et al. \cite{papernot2016distillation} as a defense against adversarial perturbations, which focused on hiding the gradients between the pre-softmax and the softmax output to provide robustness against gradient-based attacks on DNNs. This defense, however, was breached by Carlini et al. \cite{carlini2017towards} who proposed adversarial perturbation techniques that successfully evaded defensive distillation based DNNs. 

Even though the onerous job of thwarting attacks currently appears to be a Sisyphean task with no end in sight---e.g., although one can use adversarial training to train a DNN, this is a one-step solution since further adversarial examples can still be constructed for the new DNN model starting a cat and mouse game---the realization of the cognitive networking vision requires and should motivate the development of robust ML solutions.

\subsection{The Adversarial ML Challenge for Cognitive Networks}

Adversarial ML attacks have not yet been explored thoroughly for cognitive networking, although a few works have highlighted the adversarial ML threat for cognitive networks especially the real-time network telemetry component of self-driving cognitive networks. In this paper, we focus on the challenge posed by adversarial ML to the security of cognitive networking applications such as network traffic classification systems and automatic intrusion detection. 

Although numerous security attacks have been demonstrated on intrusion detection systems (IDS) \cite{corona2013adversarial}\footnote{Corona et al. \cite{corona2013adversarial} provides a detailed survey and categorizes the security attacks on IDS into six categories: evasion, overstimulation, poisoning, denial of service, response hijacking, and reverse engineering.}, little attention has focused on applying adversarial ML attacks on IDS. Similarly, there does not exist much literature on adversarial ML attacks on network traffic classification another major component of real-time network telemetry. Ahmed et al. \cite{ahmed2017poster} highlighted the problem of adversarial ML attacks on network traffic classification where they have launched an adversarial ML attack on support vector machine (SVM) based network traffic classifier and showed that a smaller perturbation in the test example can successfully evade the classifier's decision boundary and compromises the integrity of the classifier.

In our previous work \cite{usama2018adv}, we performed FGSM, BIM, and JSMA attacks on malware classifier to highlight that malware classification in cognitive self-organizing networks is extremely vulnerable to adversarial ML attacks. It has been shown in previous work that nominal feature perturbations are sufficient to fool a DNN that was previously classifying malware with $97\%$ accuracy with $0.85$ probability \cite{vorobeychik2018adversarial}. 

\section{Case Studies: Adversarial ML Attack on Intrusion Detection and Network Traffic Classification Systems}
\label{sec:caseStudy}

In this section, we present a concrete adversarial ML attack that is specific to networking applications. Instead of focusing broadly on the expansive functional area of real-time telemetry of cognitive self-driving networking, we limit our focus to using ML for two surrogate real-time telemetry cognitive networking problems: (1) anomaly-based intrusion detection, and (2) network traffic classification. The purpose of these case studies is to highlight the ease with which an adversarial ML attack can be launched and to show that many cognitive networking based ML applications in their current form may not provide any robustness against adversarial perturbations. While our explicit focus is on IDS and network traffic classification applications, our insights apply more broadly to diverse supervised, unsupervised, and reinforcement learning techniques.

We formulated the network anomaly-based intrusion detection problem as a binary classification problem, where the classification is performed between two classes; namely, ``\textit{Normal}'' or ``\textit{DoS}'' (denial of services). SVM and DNN are employed for performing the classification task. The reason for selecting SVM and DNN to perform classification is to highlight the fact that both traditional and more recent ML techniques---SVM and DNN, respectively---are highly susceptible to small carefully-crafted adversarial examples.

For the network traffic classification, we formulated it as a multi-class classification problem, where the classification is performed between ten network traffic classes; namely, \textit{WWW, MAIL, BULK, SERV, DB, INT, P2P, ATTACK, MMEDIA}, and \textit{GAMES}. We employed SVM and DNN for performing the classification task.  

\begin{figure}[h] 
    \centering    
    \centerline{\includegraphics[width=0.85\textwidth]{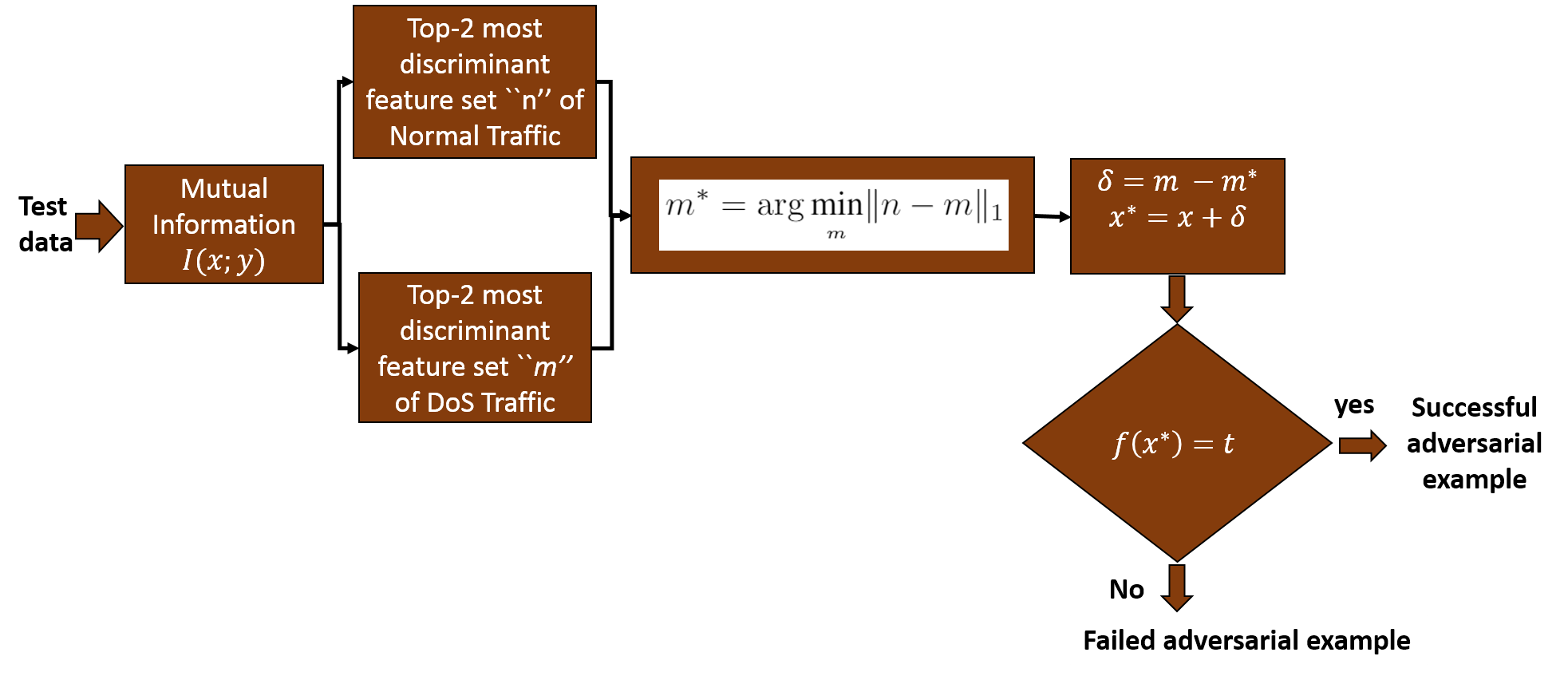}}
    \caption{Adversarial sample crafting technique where we assume white-box settings and adversarial control over test data. We employ mutual information $I(x; y)$ for extracting discriminating features and minimize $L{_1}$ norm between most discriminating features of normal and DoS classes to create an adversarial perturbation for the DoS traffic.}
    \label{fig:algorithm}
\end{figure}

\subsection{Threat Model}

\subsubsection{Adversary Knowledge}
For both case studies, we only consider evasion attacks on ML classifiers with white-box settings, where by definition the adversary has complete knowledge about the classifier's architecture, hyperparameters, and test data. We trained an SVM classifier with the radial basis function (RBF) kernel and utilized stochastic gradient descent for learning the parameters of the DNN.

\subsubsection{Adversary Goal}
We assume that the adversary wants to compromise the integrity and availability of the deployed ML-based intrusion detection and traffic classification systems. \textit{For the IDS case study}, the adversary perturbs the anomalous traffic (i.e., DoS class) while ensuring the functional behavior in such a way that the classifier mistakes it as normal traffic class. \textit{For the traffic classification case study}, the goal of the adversary is to perturb the MAIL traffic in such a way that the classifier misclassifies it in any other traffic class. Although we used the MAIL class to perform the adversarial ML attack, the proposed attack works equally well for any other target class in the dataset.

\subsection{Adversarial Sample Crafting} 
\label{adv}

\textit{For the IDS case study}, we employed the concept of mutual information $I(x; y)$ (a measure of the statistical dependence between two random variables) to find the most discriminant features in both the normal and the DoS classes. Once the most discriminant features are identified, we reduce the distance between them by using constrained $L{_1}$ norm minimization on the discriminant feature set of the DoS traffic to form a perturbation ($\delta$). The calculated perturbation ($\delta$) is then added to a DoS test example $x$ to create an adversarial DoS sample $x^*$. When the adversarial sample $x^*$ is subjected to the trained classifier $f(.)$ (which was previously classifying correctly), the classifier classifies the DoS adversarial example in the normal traffic class. Figure \ref{fig:algorithm} illustrates the steps of the proposed adversarial example generation technique. \textit{For the traffic classification case study}, we followed a similar procedure as shown in Figure \ref{fig:algorithm}, where we created adversarial examples for MAIL class. 

\subsection{Experimental Performance Evaluation}
\subsubsection{Adversarial ML Attack on IDS}
\begin{figure}[t]
\centering
\subfigure[SVM classifiers for IDS]{\includegraphics[width=0.45\linewidth]{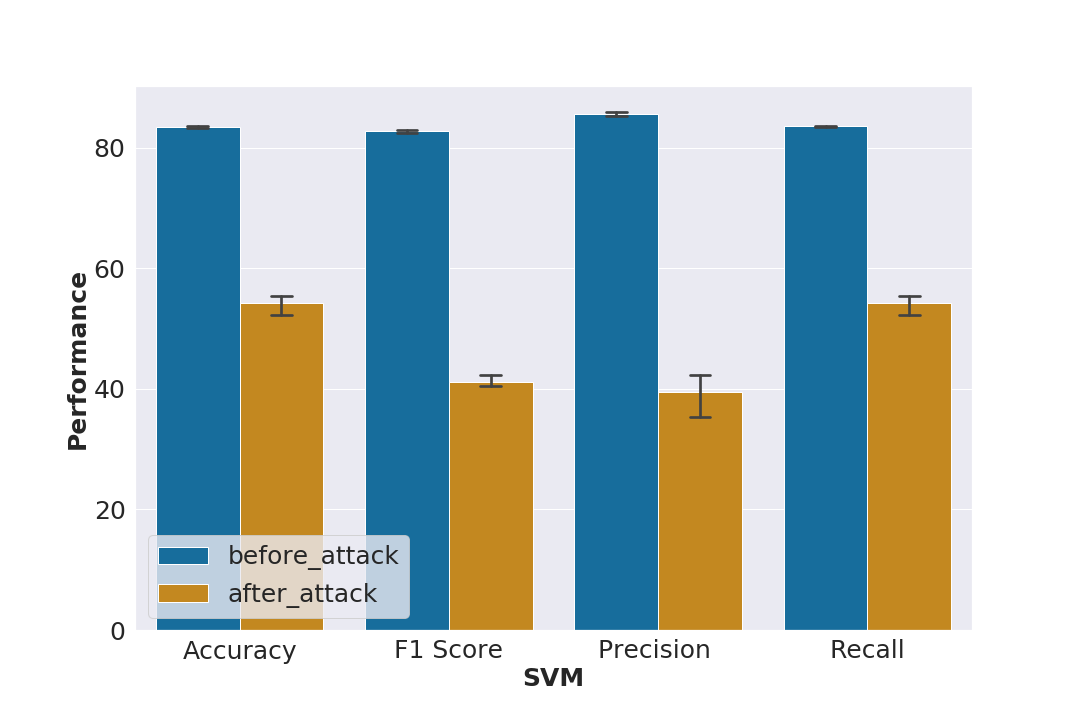}}\label{fig6}
\subfigure[SVM classifiers for network traffic classification]{\includegraphics[width=0.45\linewidth]{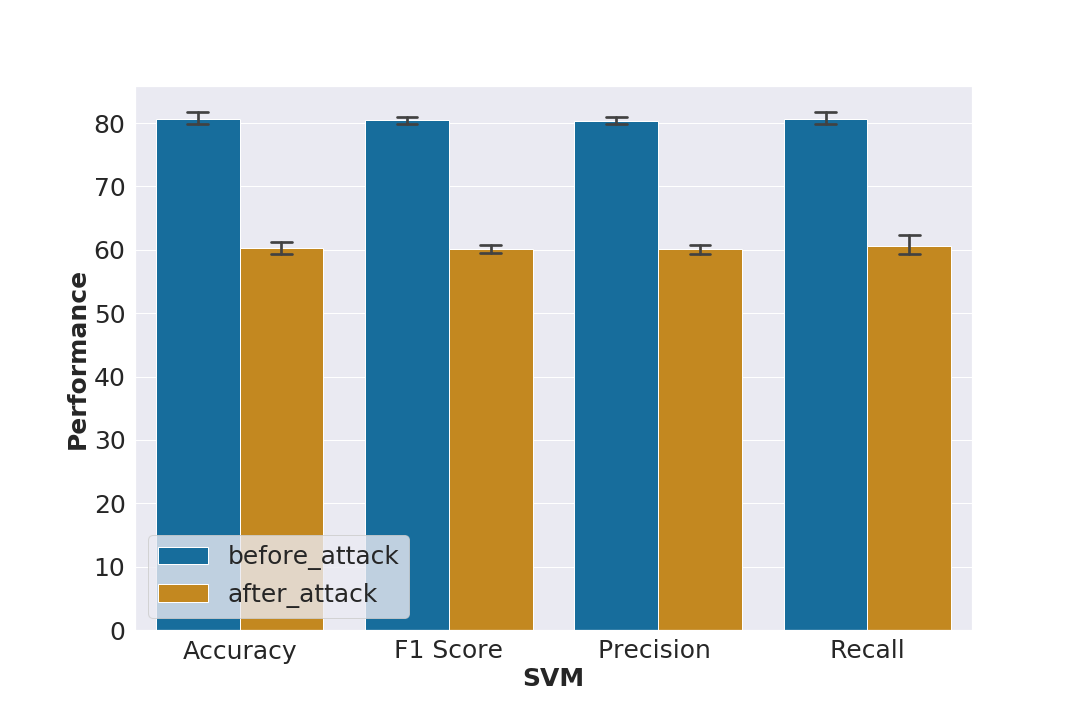}}\label{fig7}
\subfigure[DNN classifiers for IDS]{\includegraphics[width=0.45\linewidth]{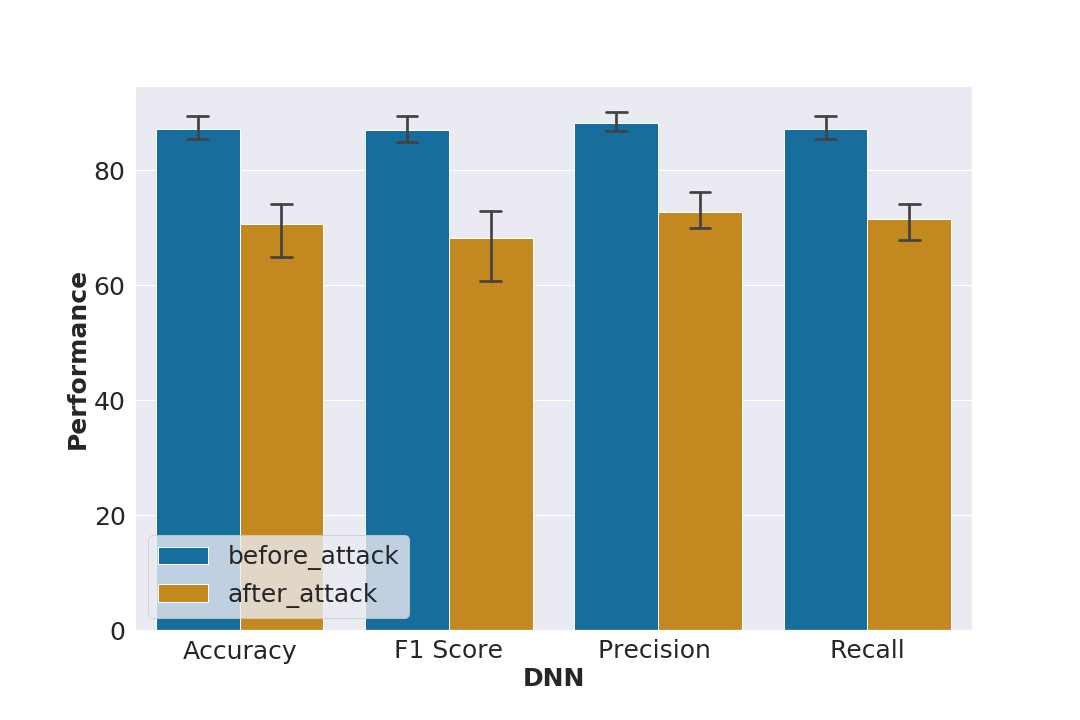}}\label{fig8}
\subfigure[DNN classifiers for network traffic classification]{\includegraphics[width=0.45\linewidth]{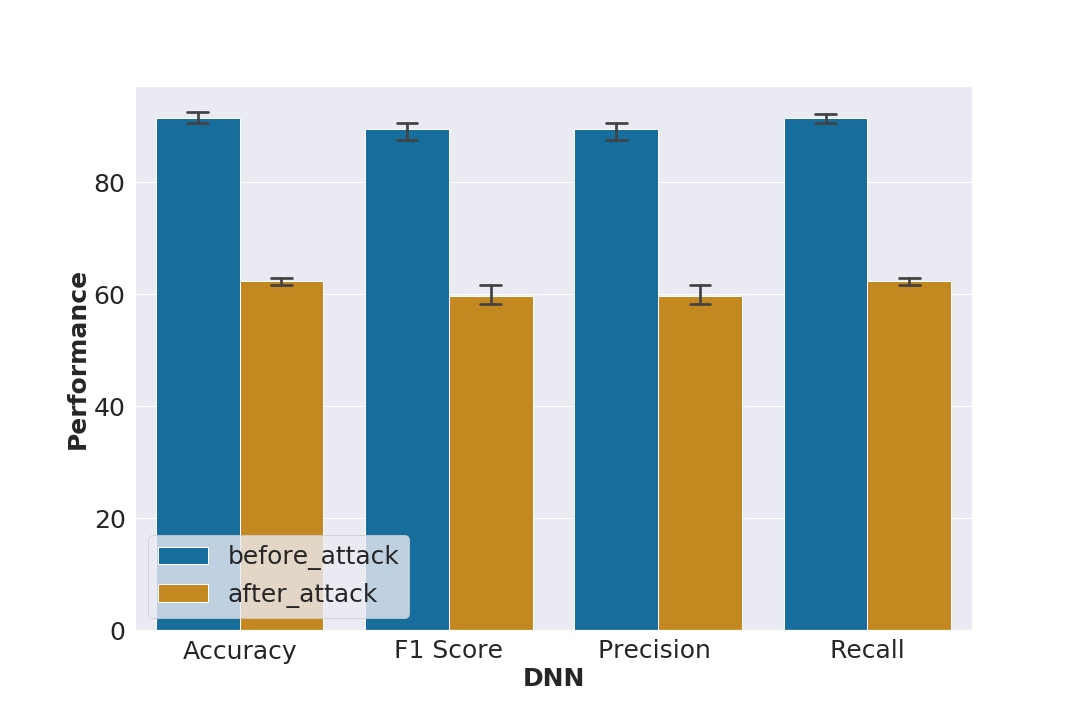}}\label{fig9}
\caption{Performance of IDS and network traffic classification before and after adversarial ML attacks.}
\label{fig5}
\end{figure}

To evaluate the performance of proposed adversarial ML attack on IDS classifier, we used the NSL-KDD intrusion detection dataset\footnote{\url{http://www.unb.ca/cic/datasets/nsl.html}}, we extracted only two classes ``Normal'' and ``DoS'' for performing this experiment. After the preprocessing, 118 traffic features were extracted in total to train the SVM and the DNN classifiers. Once the classifiers are trained, we launched an adversarial ML attack in which we generated 7460 adversarial examples (43.44\% of the complete test data) for the DoS class by perturbing \textit{only 2} out of the 118 traffic features. The size of the perturbation was constrained to be less than $10^{-2}$ to ensure the functional behavior of the DoS traffic samples. Figure \ref{fig5}(a) provides a description of various performance measures (such as accuracy, F1 score, recall, and precision) of the SVM and the DNN classifiers before and after the attack on IDS classifier.  

The proposed adversarial attack completely fooled the SVM classifier as its DoS class classification accuracy went below 1\%---rest of the adversarial samples were classified as false positives in ``Normal'' traffic category. In the case of a DNN-based intrusion detection classifier, the proposed attack successfully evaded the DNN classifier. The DoS class classification accuracy of DNN faced a 70.7\% drop in accuracy (the accuracy deterioration would have been more devastating if the number of modified features was increased). This decay in performance of DNN highlights that a very small carefully crafted input can lead DNN to a very serious malfunction. These huge drops in the performance of SVM and DNN classifiers highlight the security risk that adversarial ML poses to these ML-based techniques in cognitive networking applications.


\subsubsection{Adversarial ML Attack on Traffic Classification}

We also evaluated the performance of the proposed adversarial ML attack on network traffic classification. We used the highly cited Internet traffic classification dataset by Moore et al. \cite{moore2005internet}. The dataset consists of 377526 traffic flows divided into ten classes; namely, \textit{WWW, MAIL, BULK, SERV, DB, INT, P2P, ATTACK, MMEDIA}, and \textit{GAMES}. Further details about the dataset is provided in \cite{moore2005internet}. We deployed SVM (RBF kernel) and DNN for the classification task and achieved 89\% and 98\% classification accuracy, respectively. We used 80\% of the traffic for training the classifiers and 20\% samples for testing the performance of the classifiers. For DNN, we used four dense layers with 100 neurons per layer with ReLU as an activation function. Since we trained DNN for 10-class classification, we used softmax as an activation function in the last layer to obtain classification probabilities. We used categorical cross-entropy as a loss function to train the DNN with stochastic gradient descent (SGD) as an optimizer.  


Adversarial examples are generated for the MAIL class which is classified by the SVM classifier with 85\% accuracy and DNN classifier with 96\% accuracy. The total number of MAIL traffic samples in the test set is 5700 and we produced 700 adversarial examples by following the procedure provided in Figure \ref{fig:algorithm}. The adversarial samples successfully evaded the SVM and DNN based classifiers, where for SVM the classification accuracy of the MAIL class has fallen from 85\% to nearly 24\%. For DNN, the classification accuracy of the MAIL class has dropped from 96\% to 11\%. This drop in performance clearly highlights that the ``real-time telemetry'' component of the cognitive self-driving network is highly vulnerable to adversarial ML attacks. Figure \ref{fig5}(b) depicts the performance drop in the 10-class classification performance of SVM and DNN. 


\section{Discussion}
\label{sec:Discussion}
\subsection{Developing Robust-by-Design ML for Cognitive Networks}

It is important for ML algorithms used for mission-critical applications in cognitive networking to be robust and resilient. ML researchers in other application domains have started to work on robust-by-design models and algorithms and we should have similar, if not higher, standards for cognitive networking applications. There does not exist much work on guidelines for evaluating the defenses against adversarial ML attacks particularly against cognitive networks. In the following, we provide some guidelines leveraging the insights shared by Carlini et al. \cite{carlini2017towards} on how to effectively evaluate the robustness of ML schemes. 


\begin{enumerate}
    
    \item Check the adversarial threat model used in the defense under review for evaluating what kind of knowledge the adversary has about the targeted ML model? 
    
    \item Does the defense under review consider the presence of an adaptive adversary in the self-driving cognitive networking environment?
    
    \item Does the defense under review provide robustness against gradient-based adversarial attacks?
    
    \item Evaluate the defense under consideration using different threat model assumptions and for different performance metrics.
    
    \item Evaluate the defense under consideration against (a) strong adversarial attacks (i.e., optimization-based attacks) and not against weak attacks; (b) out of distribution adversarial examples; (c) transferable adversarial examples to check whether the transferability property of adversarial examples is blocked or not? 
    
\end{enumerate}

\subsection{Developing New Metrics for ML in Cognitive Networks}

Traditionally, the metric used to evaluate the performance of an ML model nearly always has been a variant of the metric of \textit{prediction accuracy}---i.e., how often is the model correct in its prediction or classification? To be sure accuracy can be measured in various ways (such as precision, specificity, sensitivity, recall), but using accuracy alone as a metric can only inform us of the average-case performance. This has the implicit assumption that the distribution of the test data will be similar to the distribution of the training data. This assumption obviously fails to hold when an adversary intentionally changes the test data with the explicit goal of defeating the ML model. There is, therefore, a need to also focus on evaluating a system's worst-case performance and measure the adversarial resilience of the ML model. We should move away from only using traditional ML metrics related to accuracy and precision towards a greater emphasis on robustness, transparency, and resilience. Here we recommend some new metrics for ensuring the appropriate application and validation of ML schemes in self-driving cognitive networking. There can be more metrics depending upon the design of ML-based self-driving cognitive networking application. 

\begin{enumerate}
    \item \textit{Inference Stability:} It is a measure of comparing the outputs of the victim model before and after the adversarial attack. Inference stability is calculated by measuring the distribution similarity before and after adversarial attack and defense. Divergence to the average is a popular way of computing the similarity between distributions. For self-driving cognitive networks, inference stability will provide information about the attack and recovery of the system from the adversarial ML attack.
    
    \item \textit{Classification Accuracy Variance:} It is a measure of the variance in the accuracy of the ML model before and after the adversarial attack. This metric provides the damage and recovery report after the adversarial ML attack. 
    
    \item \textit{Misclassification Ratio:} It is a measure of how many adversarial examples are classified in arbitrary classes.
\end{enumerate}

\subsection{Semantic Insights}

Despite its success in other domains, ML has traditionally not been as successful in terms of deployments in the real world for detecting anomalies. One important reason behind this is that for anomaly detection, semantic knowledge underlying the prediction, and not only the prediction itself, is important for operational networks as highlighted by Sommer and Paxson in \cite{sommer2010outside} who emphasized the need of interpretability of ML in the context of intrusion detection and anomaly detection systems in general. 
\subsection{The Broader Challenge for Adversarial ML for Cognitive Networks}
In this paper, we highlighted the threat of adversarial examples on the ML-based real-time network telemetry component of self-driving cognitive networks. Real-time network telemetry consists of supervised ML and feature engineering but there are more complex tasks in the self-driving cognitive networks (i.e., data-driven analysis and decision making). In order to perform these tasks, the network must have the ability to interact and adapt according to network conditions \cite{feamster2017and}. Deep reinforcement learning (DRL) provides the ability to interact, learn, and adapt to the ever-changing network conditions and it is expected to be heavily utilized in future self-driving cognitive networks. Unfortunately, DRL also lacks robustness against adversarial examples and it has been recently shown \cite{biggio2018wild} that adversarial examples can affect the performance of DRL. The threat of adversarial examples and brittleness of the known defenses is one of the major hurdles in the progress of cognitive self-driving networks.

\section{Conclusions}

In this article, we introduced the problem of adversarial machine learning (ML) attacks on the ML models used in cognitive self-driving networks. After introducing adversarial ML attacks, we developed novel networking-specific attacks on two surrogate real-time telemetry problems of self-driving cognitive networks to highlight their vulnerability to adversarial examples. This vulnerability to adversarial ML attacks may turn out to be the Achilles' heel of cognitive networks unless the networking and the ML communities get together to inoculate future cognitive networking from the malaise of adversarial ML attacks. The development of effective defenses against adversarial ML attacks, even though tough, is not impossible since attackers are often constrained in how effectively they can attack a model and there has been some positive progress on this front. This gives us guarded optimism that we may finally be able to develop a future of robust, resilient, and dependable ML-based cognitive networking. 







\end{document}